\documentclass{IEEE/IEEEtran}
\usepackage{amsmath,amssymb}
\usepackage{acronym,etoolbox}
\usepackage{graphicx,epstopdf}
\usepackage[labelformat=simple]{subcaption}

\usepackage{cite}
\usepackage{todonotes}

\graphicspath{{./Figures/}}
\DeclareGraphicsExtensions{{.eps}}

\usepackage{tikz}
\usepackage{xcolor}
\usetikzlibrary{decorations.pathreplacing}

\title{Distributed Coordinated Transmission with Forward-Backward Training for 5G Radio Access}

\author{
	\IEEEauthorblockN{Antti T\"olli\IEEEauthorrefmark{1}, Hadi Ghauch\IEEEauthorrefmark{2}, Jarkko Kaleva\IEEEauthorrefmark{1}, Petri Komulainen\IEEEauthorrefmark{3}, Mats Bengtsson\IEEEauthorrefmark{2}, Mikael Skoglund\IEEEauthorrefmark{2}, Michael Honig\IEEEauthorrefmark{4}, Eeva L\"ahetkangas\IEEEauthorrefmark{7}, Esa Tiirola\IEEEauthorrefmark{7} and Kari Pajukoski\IEEEauthorrefmark{7}
	} \\
	\IEEEauthorblockA{\IEEEauthorrefmark{1}\small{Centre for Wireless Communications, University of Oulu, Oulu, Finland. Email: firstname.lastname@oulu.fi}}\\
	\IEEEauthorblockA{\IEEEauthorrefmark{2}\small{Royal Institute of Technology (KTH), Stockholm, Sweden. Email: firstname.lastname@ee.kth.se}}\\
	\IEEEauthorblockA{\IEEEauthorrefmark{3}\small{Mediatek Wireless, Oulu, Finland. Email: petri.komulainen@mediatek.com}}\\
	\IEEEauthorblockA{\IEEEauthorrefmark{4} \small{Northwestern University, Evanston, IL, USA. Email: mh@eecs.northwestern.edu}}\\
	\IEEEauthorblockA{\IEEEauthorrefmark{7} \small{Nokia Bell Labs, Oulu, Finland. Email: firstname.lastname@nokia-bell-labs.com\vspace{-0.5cm}}}

\thanks{This research was supported by Academy of Finland (Decision no. 279101 and 284590), Business Finland and Nokia. Part of this work has been performed in the framework of the FP7 project ICT-317669 METIS funded by the European Union.}}

\begin{document}
	
\acused{MU}
	
\maketitle
	
\acrodef{MSE}{mean squared error}
\acrodef{IBC}{interference broadcast channel}
\acrodef{MC}{multi-cell}
\acrodef{BS}{base station}
\acrodef{MIMO}{multiple-input multiple-output}
\acrodef{SISO}{single-input single-output}
\acrodef{MU}{multiple users}
\acrodef{OFDM}{orthogonal frequency division multiplexing}
\acrodef{WSRM}{weighted sum rate maximization}
\acrodef{QoS}{quality of service}
\acrodef{SCA}{successive convex approximation}
\acrodef{SNR}{signal-to-noise ratio}
\acrodef{MMSE}{minimum \acl{MSE}}
\acrodef{SIR}{signal-to-interference ratio}
\acrodef{SINR}{signal-to-interference-plus-noise ratio}
\acrodef{Q-WSRM}{queue \acl{WSRM}}
\acrodef{QM}{queue minimizing}
\acrodef{SRA}{spatial resource allocation}
\acrodef{JSFRA}{joint space-frequency resource allocation}
\acrodef{WMMSE}{weighted \acl{MMSE}}
\acrodef{KKT}{Karush-Kuhn-Tucker}
\acrodef{GP}{geometric programming}
\acrodef{SOC}{second-order cone}
\acrodef{ADMM}{alternating directions method of multipliers}
\acrodef{PD}{primal decomposition}
\acrodef{DD}{dual decomposition}
\acrodef{FFR}{fractional frequency reuse}
\acrodef{DC}{difference of convex}
\acrodef{Q-WSRME}{\ac{Q-WSRM} extended}
\acrodef{TDD}{time division duplexing}
\acrodef{CSI}{channel state information}
\acrodef{AO}{alternating optimization}
\acrodef{OTA}{over-the-air}
\acrodef{PL}{path loss}
\acrodef{TDM}{time division multiplexing}
\acrodef{UC}{uncoordinated}
\acrodef{SUS}{semi-orthogonal user selection}
\acrodef{BiT}{bi-directional training}
\newcommand{\mbf}[1]{\mathbf{#1}}
\newcommand{\mc}[1]{\mathcal{#1}}
\newcommand{\fall}{\forall}
\newcommand{\set}[1]{\left \lbrace #1 \right \rbrace }
\newcommand{\mvec}[2]{\mbf{#1}_{#2}}
\newcommand{\ith}[1]{{#1}\mathrm{th}}
\newcommand{\pr}[1]{{#1}^\prime}
\newcommand{\mbfa}[1]{{\boldsymbol{#1}}}
\newcommand{\herm}{\mathrm{H}}
\newcommand{\sset}[1]{\left [ #1 \right ]}
\newcommand{\rfrac}[2]{{}^{#1}/{}_{#2}}
\newcommand{\eqspace}{\IEEEeqnarraynumspace}
\newcommand{\enoise}{\widetilde{N}_0}
\newcommand{\eqsub}{\IEEEyessubnumber}
\newcommand{\eqsubn}{\IEEEyesnumber \IEEEyessubnumber*}
\newcommand{\neqsub}{\IEEEnosubnumber}
\newcommand{\review}[1]{{\color[rgb]{0.1 0.1 0.5}{#1}}}
\newcommand{\trace}{\mathrm{tr}}
\newcommand{\tran}{\mathrm{T}}
\newcommand{\R}[1]{\label{#1}\linelabel{#1}}
\newcommand{\lr}[1]{page~\pageref{#1}, line~\lineref{#1}}
\newcommand{\eqn}[1]{\(#1\)}
\newcommand{\mx}{\mbf{m}}
\newcommand{\my}{\mbf{w}}
\newcommand{\mz}{\mbfa{\gamma}}
\newcommand{\mxb}{{{\mbf{m}}}}
\newcommand{\myb}{{{\mbf{w}}}}
\newcommand{\iterate}[2]{{#1}^{(#2)}}
\newcommand{\iter}[3]{{\mbf{#1}}_{#2}^{(#3)}}
\newcommand{\ma}{\mbf{x}}

\newcommand{\siter}[3]{{{#1}}_{#2}^{(#3)}}
\newcommand{\siterate}[2]{{#1}_{#2}}

\newcommand{\varx}[1]{\eqn{#1}}
\newcommand{\varxb}[1]{\eqn{\mbf{#1}}}

\newcommand{\vary}[2]{\eqn{{#1}_{#2}}}
\newcommand{\varyb}[2]{\eqn{\mbf{#1}_{#2}}}

\newcommand{\mrm}[1]{\mathrm{#1}}
\newcommand{\ipr}[1]{\bar{#1}}

\newcommand{\ncpm}[1]{\eqn{\mc{#1}}}
\newcommand{\cpm}[2]{\eqn{\widehat{\mc{#1}}^{(#2)}}}

\newcommand{\coll}[1]{\eqn{\{\mbf{#1}\}}}
\newcommand{\colls}[2]{\eqn{\{\mbf{#1}^{(#2)}\}}}

\begin{abstract}
Coordinated multipoint (CoMP) transmission and reception have been considered in cellular networks for enabling larger coverage, improved rates, and interference mitigation. To harness the gains of coordinated beamforming, fast information exchange over a backhaul connecting the cooperating base stations (BSs) is required. In practice, the bandwidth and delay limitations of the backhaul may not be able to meet such stringent demands. These impairments motivate the study of cooperative approaches based only on local channel state information (CSI) and which require minimal or no information exchange between the BSs. To this end, several distributed approaches are introduced for coordinated beamforming (CB)-CoMP. The proposed methods rely on the channel reciprocity and iterative spatially precoded over-the-air pilot signaling. We elaborate how forward-backward (F-B) training facilitates distributed CB by allowing BSs and user equipments (UEs) to iteratively optimize their respective transmitters/receivers based on only locally measured CSI. The trade-off due to the overhead from the F-B iterations is discussed. We also consider the challenge of dynamic TDD where the UE-UE channel knowledge cannot be acquired at the BSs by exploiting channel reciprocity. Finally, standardization activities and practical requirements for enabling the proposed F-B training schemes in 5G radio access are discussed.
\end{abstract}
	
	
\section{Introduction}

The performance of mobile networks is significantly limited by inter-cell interference due to the reuse of radio resources in nearby cells. Consequently, designing advanced interference coordination techniques is of utmost importance for improving the performance of a cellular network. 
Different coordinated multipoint (CoMP) variants have been included for the downlink in the 3GPP LTE-Advanced standard such as coordinated beamforming (CB) and joint transmission (JT). 
For JT-CoMP, users in the cluster are served by all the cooperating BSs, which have access to the user data. To enable such cooperation, data and CSI for the users in the cluster have to be exchanged among the cooperating BSs and possibly a centralized processor. This imposes certain requirements on the capacity and delay of the backhaul, which may not be possible in practice. Furthermore, the deployment of small cells and ultra-dense networks in future communication systems may further degrade the backhaul quality in an area of cooperating BSs. In fact, the imperfect backhaul has been recognized as one of the key issues in fast multi-cell cooperation~\cite{3G-TSGR1_TR36874}.

\begin{figure}
	\centering
	\includegraphics[width=\linewidth]{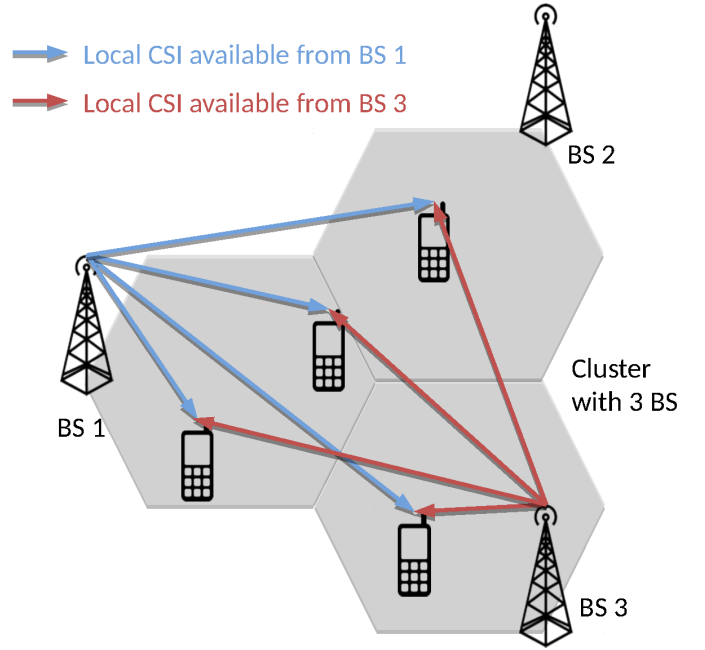}
	\caption{ Locally measured CSI.}\label{Fig:localCSI}
\end{figure}

Enabling coordination or cooperation among network entities is always beneficial but the best coordination strategy depends on a variety of conditions including channel state information (CSI) and data availability, backhaul constraints, and network connectivity (level of interference). The main design aspects are related to how much the cooperation is centrally managed (CSI sharing) and whether  terminals should be served by multiple BSs (data sharing).  
The level of coordination can be divided into four basic categories based on the different combinations of local or global CSI, and local or global data available at the network nodes.
In this paper, we focus on distributed CB-CoMP schemes~\cite{Komulainen-Tolli-Juntti-TSP-13,Kaleva-Tolli-Juntti-TSP16,Shi-Berry-Honig-14,Ghauch_IWU_15,Ghauch_MAXSEP_16} with only local CSI and data available at each transmit node as 
depicted in Fig.~\ref{Fig:localCSI}. An extension to distributed JT-CoMP scenario with minimal central controller involvement was considered in~\cite{Kaleva-Tolli-Juntti-Berry-Honig-TSP18}, assuming the user data is shared by the cooperating BSs while the CSI is available only locally.

We present iterative distributed CB schemes
that rely on uplink-downlink (UL-DL) channel reciprocity and spatially precoded pilots~\cite{Komulainen-Tolli-Juntti-TSP-13,Kaleva-Tolli-Juntti-TSP16,Shi-Berry-Honig-14,Ghauch_IWU_15,Ghauch_MAXSEP_16}. To this end, we first elaborate on how distributed CB can be facilitated via iterative forward-backward (F-B) training allowing the nodes  (BSs and users) to optimize their transmitters (TX) and receivers (RX) in an iterative manner based on local CSI only. Furthermore, we discuss the overhead resulting from the F-B iterations as well as the effect of imperfect CSI on distributed iterative schemes. 

A flexible TDD frame structure is an essential component of the upcoming 5G standard in 3GPP \cite{3GPP-TR38.802,3GPP-TR38.211}. 
The new 5G air interface must meet the low physical layer latency requirements 
without restrictions on assigning resources to UL
or DL, or in addition, to self-backhauling or direct D2D links~\cite{Lahetkangas-etal-ICC14,METIS+D23}.  
We also consider the required signaling to enable distributed coordination via F-B training in a challenging interference scenario with a fully dynamic/flexible TDD frame structure.  
Finally, practical requirements for enabling the proposed F-B training schemes in the 3GPP new radio (NR) 5G radio access are discussed. The impact on the frame structure, UE requirements, CSI uncertainty, etc. will be assessed.  

\section{Distributed Coordination via F-B Training}
\label{sec:Distr_FB_Training}

\begin{figure}
	\centering
	\resizebox{0.95\columnwidth}{!}{%
		\begin{tikzpicture}
		\draw (0,2) -- (2,2);
		\draw (0,0) -- (2,0); 
		
		\draw [fill=darkgray] (2,0)   rectangle (2.5,2);
		\draw [fill=lightgray!60]  (2.5,0) rectangle (3.0,2);
		\node[white] at (2.30,1) {{\Huge B}};
		\node at (2.77,1) {{\Huge F}};

		\draw (3,2) -- (6,2);
		\draw (3,0) -- (6,0); 
		\node at (3.75,.90) {{\Huge $\cdots$}};
		
		\draw [fill=darkgray] (4.25,0) rectangle (1+4.5,2);
		\draw [fill=lightgray!60]  (4.75,0) rectangle (1+5.0,2);
        
		\node[white] at (4.5,1) {{\Huge B}};
		\node at (5,1) {{\Huge F}};

		\draw [fill=blue!60]  (5.25,0)  rectangle (20,2);
		\node at (8.25,1) {{\Huge Data}};
		
		\draw [fill=gray!60]  (11,0)  rectangle (20,2);
		
		\draw (20,2) -- (22,2);
		\draw (20,0) -- (22,0);
		
		\node at (1,0.90) {{\Huge $\cdots$}};
		
		\node at (15.5,1) {{\Huge Frame $n\hspace{-2mm}+\hspace{-2mm}1$}};
		
		\node at (21,0.90) {{\Huge $\cdots$}};

		\draw [
		thick,
		decoration={brace,
			amplitude=15pt,
			raise=0.25cm
		},
		decorate
		] (2,2) -- (5.25,2);
				
		\node at (3.5,3.5) {\Huge $T$ iterations};
		
		\node at (8,2.75) {\Huge Frame $n$};

		\node at (10.5,-2.25-1) {\Huge OTA Signaling};
		
		\draw[-,thick,dashed] (2,0) -- (1+0,-3-1);
		\draw[-,thick,dashed](3,0) -- (1+20,-3-1);
		
		\draw [fill=gray!80]  (1+0,-3-1)  rectangle (1+9.5,-5-1);
		\node at (1+4,-4-1) {{\Huge Backward (B) pilots}};

		\draw [fill=white] (1+9.5,-3-1)  rectangle (1+10,-5-1);
		
		\draw [fill=gray!20]  (1+10,-3-1)  rectangle (1+19.5,-5-1);
		\node at (1 + 10 + 4,-4-1) {{\Huge Forward (F) pilots}};
		
		\draw [fill=white] (1+19.5,-3-1)  rectangle (1+20,-5-1);
		\draw[-,thick,dashed] (1,-7-1) -- (1,-5-1);
		\draw[-,thick,dashed](21,-7-1) -- (21,-5-1);

		\draw [fill=white,dashed] (1+8,-7-1)  rectangle (1+9,-9-1);
		
        \node at (10.5,-6.25-1) {\Huge Processing};
        
        \shade[right color=gray!20, left color=gray!80, dashed]	(1,-7-1)  rectangle (1+20,-10-1);	
        \draw [dashed]  (1,-7-1)  rectangle (1+20,-10-1);
		\node at (16,-8-1) {{\huge Forward link CSI estimation }};
		\node at (16,-9-1) {{\huge RX beamformer computation }};        
        \node at (6,-8-1) {{\huge Backward link CSI estimation }};
        \node at (6,-9-1) {{\huge TX beamformer computation }};
				
		\end{tikzpicture}
	}
	\caption{Simplified TDD frame structure with F-B training.}
	\label{fig:frame_structure}
\end{figure}
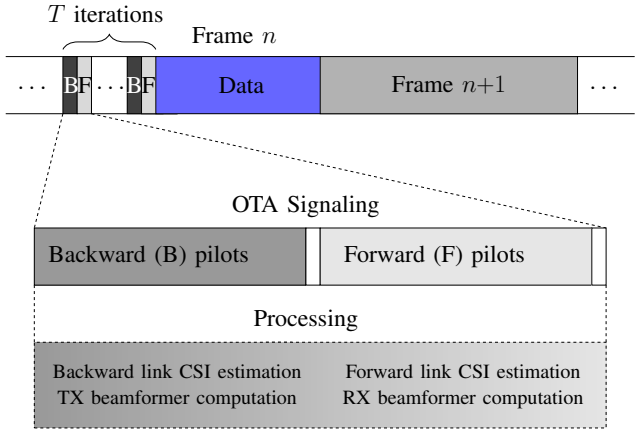

Forward-Backward iterations are often also referred to as over-the-air (OTA) iterations, or bi-directional training/signaling. A generic TDD frame structure for the proposed OTA F-B training framework is depicted in Fig.~\ref{fig:frame_structure}, where each F-B signaling round (iteration) in the beamformer training consists of two phases: a forward (F)  phase and a backward (B) phase.  Let us now focus on DL data transmission, where the BSs transmit precoded pilots in the forward phase enabling the users to estimate their local CSI consisting of the effective channel, i.e. the cascade of channel and precoder, of its desired link as well as of the interfering links. The users proceed to optimize their receive beamformers based on the acquired local CSI. The same process holds for the backward UL phase, where the transmit beamformers are computed at the BSs based on the precoded backward link pilots. Flexible UL/DL transmission is discussed in Section~\ref{sec:DynTDD}. 

F-B training allows a fully distributed coordinated computation of transmit/receive beamformers of the BSs/users in the cluster, without full CSI exchange over a backhaul. The beamformers can be computed based on different optimization criteria such as minimizing the interference leakage \cite{Ghauch_IWU_15} or the (weighted) sum MSE, or maximizing SINR~\cite{Shi-Berry-Honig-14} or the (weighted) sum rate~\cite{Shi-Razaviyayan-Luo-He-11,Komulainen-Tolli-Juntti-TSP-13,Kaleva-Tolli-Juntti-Berry-Honig-TSP18,Kaleva-Tolli-Juntti-TSP16,Ghauch_MAXSEP_16}. 
Moreover, if every step of the global optimization problem can be decoupled among the nodes, the distributed iterative scheme incurs no loss in optimality (besides the overhead) compared to the centralized approaches~\cite{Komulainen-Tolli-Juntti-TSP-13}. 

Each F-B iteration has an associated overhead due to transmitting precoded UL/DL pilots. 
A coarse measure of communication overhead can thus be given as 
the number of orthogonal pilot symbols needed for each F-B iteration. For a system with $L$ cells ($K$ users/cell), and $d$ data-streams per-user, $\Omega= T2KLd$ pilot symbols are needed, where $T$ indicates the number of F-B iterations as shown in Fig.~\ref{fig:frame_structure}, and the factor 2 follows from the even split between the number of forward and backward pilots (due to channel reciprocity)~\cite{BrandtB:16}. Thus, the minimal number of orthogonal pilots, $\Omega$, increases with the number of data streams, cells, users, and F-B iterations. In practice, assuming $\Omega$ exceeds the amount of available pilot resources, the pilots  must be reused across the network. This non-orthogonal pilot allocation causes \textit{pilot contamination} where the actual estimated channel is a superposition of all user channels reusing the same pilot resource~\cite{Jose-Ashikhmin-Marzetta-Vishwanath-2010}.


The pilot overhead $\Omega$ may become excessive for a large $T$, even with relatively low mobility, thereby destroying any potential gains from coordination. 
In this paper, the \emph{rate of the beamformer convergence} is emphasized in practical environments with time-varying channels. Faster beamformer convergence rate requires less F-B iterations, which in turn lowers the beamformer training overhead. 
The purpose of this section is to review recent work \cite{Komulainen-Tolli-Juntti-TSP-13,Kaleva-Tolli-Juntti-TSP16,Ghauch_IWU_15,Ghauch_MAXSEP_16,Shi-Berry-Honig-14}, where fast-converging F-B training and signaling algorithms are proposed (e.g, $T < 10$), based on several design criteria. 
In all papers, the general downlink CB-CoMP problem with linear TX-RX beamforming in the multiantenna interference broadcast channel (IBC) is considered. The proposed iterative methods naturally decouple the transmitter and receiver beamformer designs leading to fully distributed algorithms. Basic pilot signaling schemes for TX-RX beamformer training are introduced in~\cite{Komulainen-Tolli-Juntti-TSP-13} while alternative direct beamformer estimation approach is proposed in~\cite{Shi-Berry-Honig-14}. 
Further algorithmic convergence improvements are introduced in~\cite{Komulainen-Tolli-Juntti-TSP-13,Ghauch_IWU_15, Kaleva-Tolli-Juntti-TSP16,Ghauch_MAXSEP_16} allowing each BS/UE to quickly discard the weak streams while focusing the power to streams with large gain.

\subsection{Pilot Signalling Schemes for TX-RX Beamformer Training}
\label{sec:signalling_schemes}

In \cite{Komulainen-Tolli-Juntti-TSP-13}, 
the weighted sum rate (WSR) maximization is implemented via weighted sum mean squared error (WSMSE) minimization and alternating optimization of the transmit precoders and receivers. 
	\begin{figure}
		\centering
		\includegraphics[width=\columnwidth]{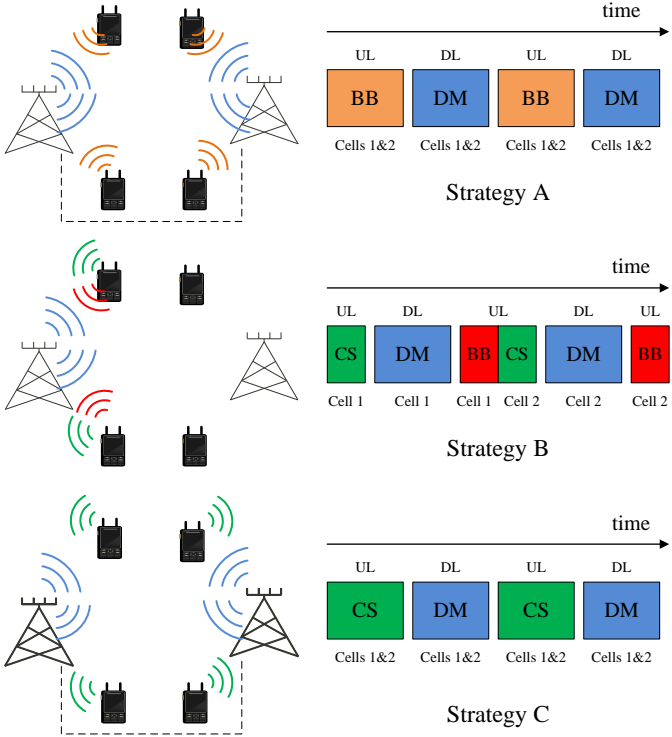}
		\caption{F-B training strategies for $T=2$ iterations with separate forward phase demodulation (DM) pilots and backward phase busy burst (BB) and whitened channel sounding (CS) pilots}
		\label{fig:signaling_AB}
	\end{figure}
%
In decentralized optimization with local CSI and data, the update of variables is distributed among the nodes in the network. 
Signaling schemes for effective CSI exchange facilitating decentralized processing are proposed in~\cite{Komulainen-Tolli-Juntti-TSP-13}, allowing the network nodes 
to locally participate in the network adaptation. Three different signaling strategies are illustrated in Fig.~\ref{fig:signaling_AB} for a special case of $T=2$ F-B iterations. The figure depicts the pilot signals employed in a network of two cells. It is worth noting that in general, the pilot signals propagate over the whole network so that all the BSs are receiving all the uplink pilots, and also all the terminals are receiving all the downlink pilots. 

%
%

In the first pilot signaling option, called \textit{Strategy A}, the TXs and the RXs are optimized consecutively using F-B training~\cite[Alg. 3]{Komulainen-Tolli-Juntti-TSP-13}. 
In the forward (F) phase, the user terminals calculate RX beamformers locally, based on the DL demodulation (DM) pilot responses. In the backward (B) phase, the resulting effective DL channels are indicated via UL pilots that are precoded by using the RX minimum mean-squared error (MMSE) filter coefficients, termed as busy bursts (BB). 
Based on the measured pilot responses, each BS locally deduces scalar weights for their own data streams, and share the weight information with other BSs through backhaul.
Alternatively, the backhaul information exchange can be avoided by incorporating the weights into additional precoded backward pilots (as done in Section~\ref{sec:DynTDD}). 
Finally, each BS optimizes its transmit precoders to be used as precoders for DM pilots in the next forward phase.  
The process is then repeated until convergence or a predefined number of F-B iteration rounds.


The idea of the second signaling option, called \textit{Strategy B}, 
is that the network-wide WSR maximization problem is split into cell-specific sub-problems, and the BSs optimize their beamformers one at a time so that each round of F-B pilot signaling involves just one cell~\cite[Alg. 4]{Komulainen-Tolli-Juntti-TSP-13}. Here,
a combination of channel sounding (CS) and BB backward phase pilots is used. The CS pilots are used to provide the effective CSI to the serving BSs such that the other-cell interference seen at the user end becomes whitened. Based on this single round of CS signals, the BS is then directly able to (iteratively) solve internally its cell-specific downlink multi-user transmit-receive design problem, which significantly reduces the amount of the required OTA signaling. 
On the other hand, similarly to Strategy A, the role of the BB pilots is to provide the adjacent BSs with the knowledge of active effective channels, concatenated with MMSE receivers and stream specific weights. 
Another benefit of Strategy B is that it requires virtually no dynamic backhaul between the BSs, as all the required information is directly exchanged via OTA signaling. 

In Strategy B, the monotonic convergence of the WSR objective is guaranteed by updating the beamformers one cell at a time.  However, when the number of coordinating BSs increases, the convergence will be slower. To speed up the process, Strategy B can be modified so that cells update their beamformers in parallel. However, this approach increases the required pilot overhead. 
To further reduce the pilot overhead of the parallel adaptation, one more approach, called \textit{Strategy C}, is proposed. 
In this scheme, the effective channels both at the serving and interfering BSs for each F-B signaling round can be constructed solely based on specifically designed whitening CS pilot responses and additional backhaul information~\cite[Alg. 5]{Komulainen-Tolli-Juntti-TSP-13}. 

\begin{figure}
	\centering
	\includegraphics[width=.5\textwidth]{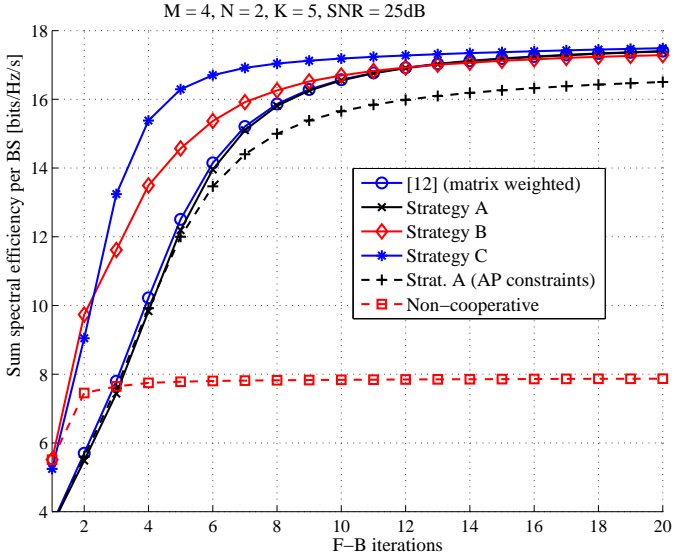}
	\caption{Average convergence of the sum rate in the cell edge at 25 dB
		SNR. Number of TX antennas $M=4$, RX antennas $N=2$, users per cell $K=5$~\cite{Komulainen-Tolli-Juntti-TSP-13}.}
	\label{fig:final_conv}
\end{figure}

Fig.~\ref{fig:final_conv} illustrates the sum spectral efficiency evolution of Strategies A--C over F-B adaptation steps, at 25 dB SNR in a 2-BS cell-edge scenario~\cite{Komulainen-Tolli-Juntti-TSP-13}. Note that SNR is representative of the total normalized power budget per BS shared among the selected multiple users/streams, and hence, the stream specific SNR can be significantly lower. \footnote{ A more thorough comparison in various operational settings can be found in~\cite{Komulainen-Tolli-Juntti-TSP-13}. The higher the cell edge SNR the larger are the relative gains from inter-cell coordination.} In addition to per-BS constraints, the performance of Strategy A with more restrictive antenna specific power (AP) constraints is also plotted in Fig.~\ref{fig:final_conv}. Here, the convergence behavior of the algorithms is emphasized while the impact of OTA signaling overhead per F-B round is considered in Section~\ref{sec:DynTDD}. 
The convergence speed is greatly improved by performing internal cell-specific iterations one BS at a time, 
as in Strategy B. 
Strategy C accelerates the convergence further as parallel iterations are allowed. 
The results demonstrate that most of the objective improvement occurs during the first 4-10 F-B iterations. 
For example, if we set a limit for F-B iterations to be $T=4$ in Fig.~\ref{fig:final_conv}, the fast converging Strategy C would give a 55\% throughput increase as compared to the baseline case. 

\subsection{Improved convergence rate of beamformer updates}

The convergence of WSMSE based algorithms~\cite{Shi-Razaviyayan-Luo-He-11,Komulainen-Tolli-Juntti-TSP-13} can be still fairly slow, especially at high SNR. In this subsection, alternative WSR algorithms with significantly faster convergence are introduced~\cite{Kaleva-Tolli-Juntti-TSP16,Ghauch_MAXSEP_16}. 
The precoded pilot signaling strategies introduced in Section~\ref{sec:signalling_schemes} can be straightforwardly incorporated into the faster converging beamformer design algorithms~\cite{Kaleva-Tolli-Juntti-TSP16,Ghauch_MAXSEP_16}.
As a result, the number of required F-B iterations $T$ in Fig.~\ref{fig:frame_structure} can be potentially reduced, further minimizing the OTA beamformer training overhead.

A WSR maximization framework based on successive convex approximation was proposed in~\cite{Kaleva-Tolli-Juntti-TSP16}
, also with additional per-user QoS/rate constraints. 
Similarly to~\cite{Shi-Razaviyayan-Luo-He-11,Komulainen-Tolli-Juntti-TSP-13}, the precoder design is based on MSE minimizing reformulation of the original WSR maximization problem, where the complexity is restricted to a set of non-convex MSE constraints. 
The framework in~\cite{Kaleva-Tolli-Juntti-TSP16} reveals a structure that enables the use of heuristic approximation methods that can be used to significantly improve the rate of convergence and lower the required number of over-the-air iterations as compared to the baseline case~\cite{Shi-Razaviyayan-Luo-He-11}. These approximation techniques are based on a prediction of the stream specific rate progression and overestimating the next point of approximation. 

%

The WSR maximization is also tackled in~\cite{Ghauch_MAXSEP_16}, where the sum-rate is lower bounded using a \emph{Difference of Log and Trace (DLT) bound}, and its relative tightness is established.
Similarly to  weighted MSE reformulation approaches used in~\cite{Shi-Razaviyayan-Luo-He-11,Komulainen-Tolli-Juntti-TSP-13, Kaleva-Tolli-Juntti-TSP16}, the DLT bound naturally decouples at the transmitters and receivers, thus leading to fully distributed algorithms. 
The proposed method has an inherent feature to switch off streams exhibiting low SINR levels, thereby significantly accelerating the algorithm's convergence. 
The DLT algorithm is  benchmarked in~\cite{Ghauch_MAXSEP_16}  against several known schemes, e.g., distributed Interference Alignment (IA), max-SINR, weighted MMSE, in terms of performance and overhead, where the fast-converging nature of the proposed algorithms is made clear: more than $95\%$ of the final performance is reached in just $2$ iterations, allowing $2$-$3$ times faster convergence (and a corresponding reduction in overhead). 
We refer the readers to~\cite{Ghauch_MAXSEP_16} for extensive simulation results.

Apart from WSR maximization, the proposed F-B training framework can be adopted to any network optimization objective solved via iterative TX-RX beamformer optimization.
For example, the interference leakage utility is tackled in~\cite{Ghauch_IWU_15}, by relaxing the constraint on the well-known leakage minimization problem. 
A rank-reducing filter update structure is introduced, to solve the resulting non-convex problem, and gradually reduce the BS/UE filter rank, thus decreasing the dimension of the interference subspace, and greatly speeding up the convergence. 
Similarly to Strategy B/C introduced in Section~\ref{sec:signalling_schemes},
a \emph{turbo-like} structure is also proposed {for the interference leakage minimization} in~\cite{Ghauch_IWU_15}, where, in addition to the outer F-B iteration loop, a separate inner optimization is carried out at each receiver. 
The proposed algorithms are shown to converge to locally optimal solutions with drastically reduced F-B iterations as compared to the hundreds of iterations required by conventional distributed interference alignment algorithms~\cite[Fig. 2]{Ghauch_IWU_15}. 
Moreover, the performance gap between proposed and conventional algorithms increases with the system dimensions (e.g., antennas, users, cells). 
We refer interested readers to extensive numerical results in \cite{Ghauch_IWU_15}.

\subsection{Impact of non-orthogonal pilots and direct filter estimation}
Pilot contamination or non-orthogonality has been widely studied in the literature~\cite{Jose-Ashikhmin-Marzetta-Vishwanath-2010}. More generally, the impact of imperfect CSI has been a popular topic. Many different strategies can be developed for estimating and sharing effective CSI and/or directly estimating beamformers from pilots. Some of these possibilities are investigated in \cite{BrandtB:16}, providing different trade-off among pilot overhead, CSI sharing over the air and via backhaul, computational complexity and resulting performance. Also, the importance of adding robustness to estimation errors is highlighted. 

An alternative approach to joint optimization of TX and RX beamformers is proposed in~\cite{Shi-Berry-Honig-14}.  There the beamformers are estimated {\em directly} as adaptive filters, e.g., using a least squares criterion. All UL/DL pilots must then be transmitted {\em synchronously} across the cells. That is, the UL pilots are transmitted synchronously from the mobiles to estimate the TX beamformers, while the RX beamformers are estimated directly from the synchronous DL pilots. 
This scheme automatically suppresses both intra- and inter-cell interference, in the MMSE sense given sufficient training, and is well-suited to the scenario in which each BS must compute its beamformers with limited or no information exchange with neighboring BSs. The pilot synchronization requirement effectively replaces the pilot coordination problem in CSI-based estimation. That is, BSs can choose the pilot sequences independently without knowledge of the neighboring cell pilots. As with CSI-based estimation, orthogonalizing and jointly designing the pilot sequences can reduce the estimation error given a fixed training duration. 

The direct beamformer estimation approach potentially reduces the pilot overhead in dense systems, relative to estimating CSI. This is due to the large number of cross-channels causing significant inter-cell interference. All those channels must be estimated when computing the beamformers, requiring knowledge of all interfering pilot sequences. In contrast, the direct beamformer-estimation approach requires only local pilot sequence information and sufficient pilots to estimate the beamformer (not channel) coefficients. In~\cite{Kaleva-Tolli-Juntti-Berry-Honig-TSP18}, a direct beamformer estimation method combined with F-B OTA training was developed for distributed JT-CoMP operation, where the user data is presumed to be shared by the cooperating BSs while the CSI is available only locally.

\section{Dynamic/Flexible TDD}
\label{sec:DynTDD}

In small cell environments, the instantaneous traffic in UL and DL may exhibit significant variation with time and among the neighboring cells. 
In such a scenario, \textit{Dynamic or Flexible TDD} may provide vastly enhanced resource utilization by dynamically changing the amount of resources allocated to the UL and DL at each time instant. 
In addition to the traditional DL-to-DL and UL-to-UL   interference, UL-to-DL and DL-to-UL interference are also associated with flexible TDD operation. 
Effective resource allocation in such demanding environment requires that the gains of flexible UL/DL allocation (e.g. packet delivery time reduction) and the potential losses (excess interference) are properly balanced. 

In order to mitigate the cross-link interference, joint resource allocation and beamforming may be employed within a coordinating set of cells.
In TDD networks, the CSI of the BS-user and BS-BS links can be acquired assuming the UL-DL channel reciprocity holds. 
A specific challenge with the dynamic TDD setting is the CSI acquisition of the cross-link interference channels (among mutually interfering user terminals). Explicit feedback of the user-to-user channels in addition to a full CSI exchange between BSs would be required to enable optimal beamformer design. 
However, such an impractical centralized approach can be circumvented by employing F-B training to convey implicit information to the BSs about the interference and receivers at the users~\cite{Jayasinghe-Tolli-Latva-aho-SPAWC15}. A flexible frame structure similar to Fig.~\ref{fig:frame_structure} can be used for the bi-directional beamformer training, where the 
data payload can be allocated either for DL or UL depending on the instantaneous traffic requirements of a given cell. 

\begin{figure}
	\centering
	\includegraphics[width=0.5\textwidth]{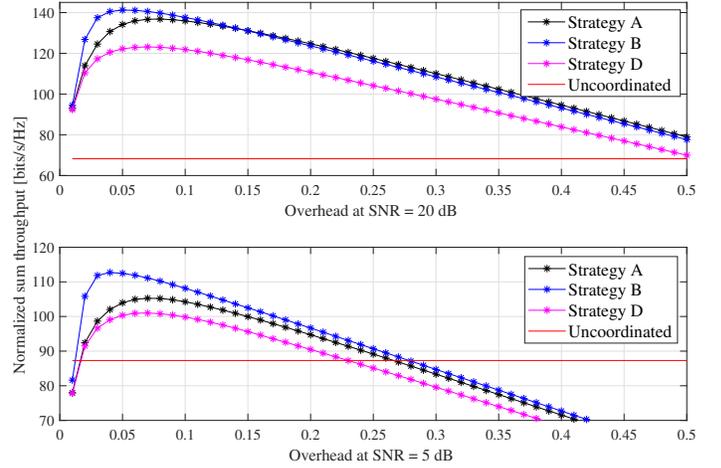}
	\caption{Sum throughput performance vs training overhead of distributed CB in a 19-cell dynamic TDD network with 20 dB and 5 dB cell edge SNR. The probability for a cell to be in UL mode is 30\%.}
	\label{fig:overhead}
\end{figure}

In Fig.~\ref{fig:overhead}, the performance of distributed CB with dynamic TDD is evaluated in a  2-tier cellular system with a wrap-around hexagonal grid consisting of 19 cells with 200m inter-site distance. Each cell is randomly allocated to operate either in UL or DL. Each BS has 8 antennas and serves four 2-antenna users in its cell. 
Two bi-directional F-B signaling strategies (A and B) introduced in Section~\ref{sec:signalling_schemes} are investigated. The forward (F) pilots correspond to precoded pilots transmitted by UL users and DL BSs while in the backward (B) phase precoded pilots are transmitted by UL BSs and DL users.  
In Dynamic TDD operation, both users and BSs participate in the iterative beamformer optimization process. Hence, \textit{wired backhaul is not available for exchanging weight information among mutually interfering UL and DL users}. However, the stream specific weights can be incorporated into additional backward pilots as illustrated in~\cite[Fig. 1]{Jayasinghe-Tolli-Latva-aho-SPAWC15}. Thus, both Strategies A and B introduced in Section~\ref{sec:signalling_schemes} necessitate two precoded pilots for each backward iteration. 
As a low overhead alternative solution, a heuristic scheme denoted as Strategy D is also proposed. The weight variables are assumed to be all ones for the other cell users, while the MSE weights for the own cell users are calculated based on locally available information~\cite{Jayasinghe-Tolli-Latva-aho-SPAWC15}. Thus, only a single backward pilot per stream is required in Strategy D.

The effective sum throughput versus the total overhead 
is shown in Fig.~\ref{fig:overhead} for each proposed training strategy. 
The effective sum throughput is given as  $(1-T\gamma)R$, where $R$ is the achieved sum rate  from the iterative algorithm after $T$ F-B iterations and $\gamma$ denotes the relative overhead per one F-B signaling round.  In Fig.~\ref{fig:overhead}, 
one signaling round is assumed to use up 1\% of the available resources per scheduling interval for Strategies A and B, while the per iteration overhead of Strategy D is just 2/3 of the two other schemes. 
The numerical example in Fig.~\ref{fig:overhead} point out an apparent trade-off between the number of F-B iterations available and the improvement of effective throughput due to iterative beamformer updates. 

All F-B strategies provide significant gain (up to 110\% at high SNR) as compared to the uncoordinated scenario where the transmit beamformers are designed locally ignoring the inter-cell interference altogether. In this particular example, about 4-8\% of the resources, depending on the selected strategy, should be allocated to the F-B training in order to maximize the total throughput. Moreover, Strategy B provides a significant additional gain since internal iterations are allowed at the BS  between each F-B round. 
Despite higher backward phase training overhead per iteration, both strategies A and B provide better peak throughput than Strategy D due to optimized MSE weight variable adjustments among coordinating cells.


\section{Standardization Activities and Practical Aspects}
\label{sec:practical_aspects}

\begin{figure}
	\centering
		\includegraphics[width=.5\textwidth]{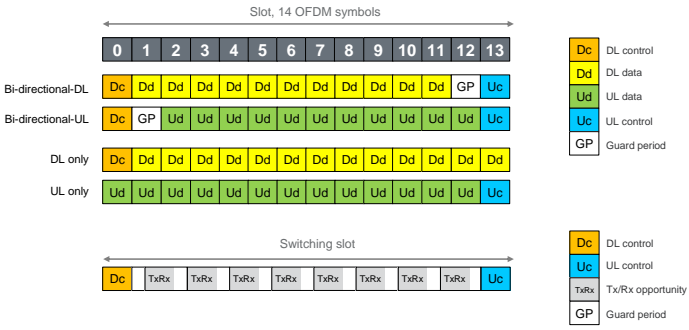}
	\caption{New radio slot types agreed in 3GPP NR and the proposed switching slot structure.}
	\label{fig:new_radio_slot_types}
\end{figure}

3GPP new radio (NR) specification~\cite{3GPP-TR38.211} will support multiple OFDM parameter sets and time-frequency scaling of LTE. 
A slot consisting of 14 OFDM symbols defines the basic scheduling interval. NR supports at least four slot types, illustrated in the upper part of Fig.~\ref{fig:new_radio_slot_types}, providing the basic support for both TDD and FDD modes: 1) bi-directional slot with DL data, 2) bi-directional slot with UL data, 3), DL only slot, 4) UL only slot. These different slot types can be concatenated and aggregated in a flexible manner. Bi-directional slot types, including a bi-directional control signal part embedded in each slot and time separated from the data payload, are required in TDD mode to facilitate link direction switching between DL and UL. They also enable fully flexible traffic adaptation between the link directions and with opportunity for low latency. Demodulation reference signal (DMRS) symbols are located e.g. in the first symbol of the data part and can be precoded with the data, enabling the receivers to estimate the equivalent channel of its desired and interfering links. 
Unlike LTE, 3GPP NR will support non-codebook based multi-stream uplink transmission. Furthermore, precoded uplink sounding reference signals are supported providing means for iterative TX-RX beamformer training.
 Since NR needs eventually to support carrier frequencies up to 100 GHz, the NR frame structure should eventually also support integrated access and backhauling scenarios for deployments with high carrier frequencies.

The aggregated scheduling block should be sufficiently long to allow for sufficient beamformer convergence and to avoid excessive overhead due to F-B training. Also, the user allocation in the adjacent cells should not change during the F-B training to allow fast adaptation to the interference scenario. Therefore, a scheduling block consisting of multiple aggregated slots would be useful in practice. However, there is a clear trade-off between the channel coherence time and the size of the scheduling interval, as well as traffic burstiness.  

It can be noted that the basic NR bi-directional slot types defined in 3GPP and illustrated in the upper part of Fig.~\ref{fig:new_radio_slot_types} cannot provide support for multiple TX/RX updates, i.e., multiple bi-directional F-B rounds within a slot. In principle, the related TX/RX switching could be distributed among consecutive slots within the scheduling block. However, in order to isolate the impact of sequential TX/RX switching operation and to minimize the involved latency, it would make sense to concentrate TX/RX switching functionality into specific switching slots. As shown in the lower part of Fig.~\ref{fig:new_radio_slot_types}, an additional switching slot type used at the beginning of the scheduling block could consist of a plurality of TX/RX opportunities and also enable multiple bi-directional F-B rounds. In such a case, the beamformer computation both at UEs and BSs must occur within each TX-RX update, which may pose challenges for practical implementation. This type of switching slot functionality may in practice be implemented via a mini-slot structure also already agreed in NR 3GPP~\cite{3GPP-TR38.211}. 
Utilization of mini-slot structure allows having more than one TX/RX (DL/UL) switching points within a 14-symbol slot by using non-slot-based scheduling. Hence, a flexible frame structure or scheduling block similar to Fig.~\ref{fig:frame_structure} can be constructed by proper concatenation of mini-slots and UL/DL slots (see Fig.~\ref{fig:new_radio_slot_types}).

In order to support F-B training functionality, the \textit{terminals should start performing similar functions as BSs have traditionally done}, i.e., being more aware of the neighborhood and measuring the other nodes (both users and BSs) in the near vicinity.  The UEs should be able to measure pilots, including the nearby users operating in reverse UL/DL mode in order to be able to compute their respective TX/RX beamformers to avoid excessive UE-UE interference.   Moreover, F-B training requires a dedicated radio frequency chain per antenna to allow for pilot precoding at both the BS and UE. Also, sufficient calibration of TX/RX radio frequency chains at both ends is required.

The concatenation of both analog and digital beamforming functions makes the \textit{mmWave communication} scenario more challenging for the F-B training based distributed coordination. Ideally, with fully digital beamforming, the signal at each antenna of the BS/UE can by digitally  controlled/processed.  
However, the hybrid analog-digital precoding, prevalent in mmWave communications due to the limited number of radio frequency chains,   
requires separate beam search mechanisms, thus complicating the implementation of additional OTA F-B training procedures. 
Note that the simulation studies in~\cite{Ghauch_MAXSEP_16} assuming all-digital beamformer implementation demonstrate that simple inter-cell interference coordination mechanisms still provide significant gains in sub-28 GHz frequencies in dense deployments.

\section{Conclusion}

Distributed schemes for the multi-cell coordination were discussed in this paper. Motivated by the fact that several deployments in future wireless communication networks might have a backhaul with limited capabilities, approaches based solely on local CSI were considered. This assumption eases the burden on the backhaul by avoiding CSI exchange between the cooperating BSs, thus addressing core aspects for small cell integration. For this purpose, distributed coordination based on F-B training was proposed to gradually refine the transmit/receive beamformers of the nodes in a fully distributed manner. Several relevant issues of such iterative schemes were addressed like the training overhead, signaling, and imperfect CSI. We have also shown how F-B training can be employed to manage interference in a dynamic TDD system. The numerical experiments demonstrated that the proposed F-B strategies achieve considerable gain in comparison to the reference case without interference coordination.  In the dynamic TDD case, for example, gains of up to 100\% can be achieved with an overhead of less than 5\% at an SNR of 20 dB. On the other hand, both UE and BS capabilities should be enhanced in order to support F-B training functionality. A new switching slot structure extension was introduced on top of the already agreed 3GPP NR slot types to support multiple bi-directional F-B rounds within a scheduling interval.



\bibliographystyle{IEEE/IEEEtran}
\bibliography{References/IEEEabrv,References/IEEEfull,References/conf_short,References/kirja_survey,References/refs_antti}

\begin{IEEEbiographynophoto}{Antti T\"olli}(M'08--SM'14) received the D.Sc. (Tech.) degree in electrical engineering from the University of Oulu, Oulu, Finland, in 2008. From 1998 to 2003 he worked with Nokia Networks both in Finland and Spain. Currently, he holds an Associate Professor position with the Centre for Wireless Communications, University of Oulu.  He has held visiting positions both at  EURECOM, France, and University of California - Santa Barbara, USA. 
\end{IEEEbiographynophoto}
\begin{IEEEbiographynophoto}{Hadi Ghauch} (S'06) is a postdoctoral researcher at school of EECS at the Royal Institute of Technology (KTH), Stockholm, where he also received his PhD in Jan 2017. Prior to that, he received his MS from Carnegie Mellon University, Pittsburgh, in 2011.  His research interests includes optimization for learning large scale learning, optimization for millimeter-wave communication, and distributed optimization of wireless networks.
\end{IEEEbiographynophoto}
\begin{IEEEbiographynophoto}{Jarkko Kaleva} received his Dr.Sc. (Tech) in telecommunications from University of Oulu, Oulu, Finland in 2018. In 2010, he joined Centre for Wireless Communications (CWC) at University of Oulu, Finland, where he is currently working as a post-doctoral researcher. His main research interests are in resource management for millimeter wave wireless communications, cross-layer optimization, coded caching and nonlinear programming. 
\end{IEEEbiographynophoto}

\begin{IEEEbiographynophoto}{Petri Komulainen}  received the D.Sc (Tech.) degree in communications engineering in 2013 from the University of Oulu, Finland. He has more than 20 years of industry and academic experience in the field of wireless communications, working on various topics around 3G/4G/5G systems research and physical layer algorithm development. He is currently with MediaTek Wireless Finland.
\end{IEEEbiographynophoto}

\begin{IEEEbiographynophoto}{Mats Bengtsson} (M'00--SM'06) received a M.S. degree in computer science from Linköping University, Sweden, in 1991 and a Ph.D. degree in electrical engineering from KTH Royal Institute of Technology, Sweden, in 2000. From 1991 to 1995, he was with Ericsson Telecom. He currently holds a position as Professor at the Information Science and Engineering department, School of EECS, KTH. His research interests include statistical signal processing and optimization theory and its applications to communications.
\end{IEEEbiographynophoto}

\begin{IEEEbiographynophoto}{Mikael Skoglund} (S'93--M'97--SM'04) received the Ph.D.~degree in 1997 	from Chalmers University of Technology, Sweden. In 1997, he joined
	the Royal Institute of Technology (KTH), Stockholm, Sweden, where he
	was appointed to the Chair in Communication Theory in 2003. At KTH,
	he heads the Department of Information Science and Engineering.	
	Dr.~Skoglund has contributed to information theory, wireless
	communications, signal processing and control. He has authored more
	than 140 journal papers in these areas.
\end{IEEEbiographynophoto}

\begin{IEEEbiographynophoto}{Michael L. Honig} (M'81--SM'92--F'97) received the B.S. degree in electrical engineering from Stanford University in 1977, and the M.S. and Ph.D. degrees in electrical engineering from U.C. Berkeley in 1978 and 1981, respectively. He subsequently joined Bell Laboratories in Holmdel, NJ, and in 1983 he joined the Systems Principles Research Division at Bellcore. Since Fall 1994 he has been a Professor in the EECS department at Northwestern University.
\end{IEEEbiographynophoto}

\begin{IEEEbiographynophoto}{Eeva L\"ahetkangas} received her M.Sc. degree in communication engineering in 2005 from the University of Oulu. She has been with Nokia since 2005, where she first concentrated on digital signal processing for WCDMA, WiMAX, and LTE. Currently she is concentrating on 5G physical layer research.
\end{IEEEbiographynophoto}

\begin{IEEEbiographynophoto}{Esa Tiirola} received his M.S.E.E. in 1998 from the University of Oulu. He is currently with Nokia Bell Labs, where he is working on various topics related to radio research and standardization. His current research interests include signal processing and physical layer design for 5G.
\end{IEEEbiographynophoto}

\begin{IEEEbiographynophoto}{Kari Pajukoski} received his B.S.E.E. degree from Oulu University of Applied Sciences in 1992. He is a Bell Labs Fellow, and has broad experience in cellular standardization, link and system simulation, and algorithm development for products. Currently he is focusing on key physical layer technologies for 5G.
\end{IEEEbiographynophoto}

\end{document}